\documentclass[aps,prb,twocolumn,showpacs]{revtex4}

\usepackage{amsmath}
\usepackage{color}
\usepackage{amssymb}

\usepackage{graphicx}

\begin{document}

\title{Magnetic relaxation  in uranium  ferromagnetic superconductors}

\author{V.P.Mineev}
\affiliation{Commissariat a l'Energie Atomique, INAC / SPSMS, 38054 Grenoble, France}

\begin{abstract}
There is  proposed a phenomenological description of quasi-elastic neutron scattering in the  ferromagnetic metals UGe$_2$ and UCoGe 
based on their property that  magnetization  supported by the moments located at uranium atoms  in an orthorhombic  crystal is not conserved quantity relaxing to equilibrium by the interaction with itinerant electron subsystem.  As result the line width of quasi elastic neutron scattering at $q\to 0$ acquires nonvanishing value at all temperatures but the Curie temperature. 
\end{abstract}

\date{\today}
\maketitle
%\section{Introduction}
The family of heavy fermionic uranium metallic compounds UGe$_2$, URhGe and UCoGe  possesses   astonishing property of coexistence of superconductivity and ferromagnetism ( for the most recent reviews see \cite{Aoki11,Aoki12}).  Ferromagnetism does not suppress the superconductivity with triplet pairing and since the discoveries of superconductivity in uranium ferromagnets they were considered as 
equal spin pairing superconductors similar to  $^3$He-A and $^3$He-A$_1$ superfluids. The pairing interaction in liquid helium is due to spin-fluctuation exchange, hence, it was quite natural  to consider the same mechanism as the origin of superconductivity in uranium compounds. There was implied that they are fully itinerant ferromagnets and the same $5f$ electrons are
responsible for ferromagnetism and superconductivity.  Thus, practically in all publications the uranium ferromagnet superconductors 
were considered  in frame of theory of the isotropic Fermi liquid with ferromagnetism induced by the Landau-Stoner interaction between electrons (one can find the list of corresponding references in review \cite{Aoki11}). This  beautiful theoretical model is reasonable for $^3$He.  But in relation  to the uranium compounds its applicability is quite doubtful in view of 
significant
 crystallographic and magnetic anisotropy, as well because of  non-itinerant nature of magnetism in these materials.

The static magnetic properties of UGe$_2$  are well described in Ref. 4 in terms of crystal field splitting of  $U^{4+}$ state, which is $^3H_4$  term of the  $5f^2$ configuration of localized electrons, despite of the presence of the itinerant electrons filling the bands formed by two $7s$, one $6d$ and one $5f$ uranium and also germanium orbitals. So, UGe$_2$  is actually dual system where local and itinerant states of $f$-electrons coexist. The example of such type coexistence has been clearly demonstrated \cite{Feyerman} by $\mu SR$ measurements in another uranium compound UPdAl$_3$ where the Knight shift below $T_{sc}$ indicates that local moment magnetism and superconductivity are carried by different electron substrates of $5f$ character, one of which involves the heavy quasiparticles.
The localized nature of magnetism in  uranium ferromagnetic compounds
put forward as  the most plausible pairing mechanism  the interaction between the conduction electrons by means of spin waves in the system of localized moments. The first such type model has been applied to the superconducting antiferromagnet UPd$_2$Al$_3$ \cite{McHale} and then quite recently to the reentrant ferromagnetic superconductor URhGe  \cite{Hattori}.
 
 Leaving for following publications the superconducting properties we  discuss here the problem of magnetic excitations.

The magnetic excitations reveal themselves in neutron scattering measurements of dynamical structure factor $$S_{\alpha\beta}({\bf q},\omega)=\int_{-\infty}^\infty dt e^{i\omega t}\langle M_{\alpha{\bf q}}(t)M_{\beta-{\bf q}}(0)\rangle_{eq}$$ which is wave vector - frequency dependent magnetic moments correlation function \cite{Hove}. For  an  isotropic ferromagnet $S_{\alpha\beta}({\bf q},\omega)=S({\bf q},\omega)\delta_{\alpha\beta}$. 

In absence of walls and spin-orbital coupling the magnetization is conserved quantity, hence, in isotropic Heisenberg ferromagnet above Curie temperature  the only mechanism  leading to the magnetization relaxation is the spin diffusion that results in \cite{Hove,Forster}
\begin{equation}
S({\bf q},\omega)=\frac{2\omega\chi({\bf q})}{1-\exp(-\frac{\omega}{T})}\frac { \Gamma_{\bf q}}{\omega^2+(\Gamma_{\bf q})^2},
\label{d}
\end{equation}
such that line width of quasi elastic scattering 
\begin{equation}
\Gamma_{\bf q}=Dq^2
\label{diff}
\end{equation}
is determined by the diffusion coefficient $D$. Here $ \chi({\bf q})=\frac{\chi_0}{1+(\xi q)^2}$,
$\chi_0$ is the static uniform susceptibility.  We put the Planck constant $\hbar=1$.
The $q^2$ law dependence was observed in wide temperature range above $T_c$ in Ni and  Fe ( see \cite{Shirane} and references therein) reducing at $T=T_c$ to $\Gamma\propto q^{2.5}$ dependence according to predictions of mode-mode coupling theory \cite{Hohenberg}.  

In weak itinerant ferromagnets above Curie temperature another mechanism of dissipationless relaxation  can dominate with structure factor given by the same Eq. (\ref{d}) 
but with 
the line width determined by equality \cite{Hertz,Moriya}
\begin{equation}
\chi({\bf q})\Gamma_{\bf q}=\chi_P\omega({\bf q}) 
\label{Landau}
\end{equation}
where $\chi_P$ is the noninteracting Pauli susceptibility, $\omega({\bf q})$  is the Landau damping frequency equal to  $\frac{2}{\pi}qv_F$ for the spherical Fermi surface.
The linear in wave vector line width was observed  in MnSi \cite{Ishikawa}, however, in the other weak itinerant ferromagnets MnP \cite{Yamada} and Ni$_3$Al 
\cite{Semadeni} the line width q-dependence is closer to the dynamic scaling theory predictions \cite{Hohenberg}.

The investigations of magnetic excitations in UGe$_2$ and UCoGe
has been reported in several publications \cite{Huxley,Raymond,Stock}. The main result is that $\Gamma_{\bf q}$ unlike both Eqs (\ref{diff}) and (\ref{Landau}) does not vanish as $q\to 0$ for temperatures different from $T_c$.\cite{residual} The authors \cite{Huxley} fairly specify
that the finite value of $\chi({\bf q})\Gamma_{\bf q}$ as $q \to 0$ implies that the uniform magnetization density is not conserved quantity. 
The only source of violation of magnetization conservation in  an itinerant electron system is electron-electron spin-orbit coupling.
Hence, for the significant non-spin-conserving mechanism they have proposed the spin-orbit interaction associated with $f$ electrons 
silently assuming that for $f$-electron system  the inter-electron spin-orbit interaction is stronger than in the ordinary metals. Indeed, the intraatomic spin-orbital coupling
is important at calculation of the electron  band structure in compounds consisting of elements with big atomic numbers. However, it is well known that electron-lattice spin-orbital interaction in the crystals with inversion center plays role similar,
or better to say, equivalent to the usual spin-independent interband transition terms leading to the additional band splitting but not eliminating the Kramers double degeneracy of electronic states. So, when the one-electron band structure is fixed 
 one can work  with electron-electron interaction independent of initial atomic orbitals used for the band construction. Hence, the electron-electron spin-orbital interaction has usual relativistic smallness $\sim(v_F/c)^2$.
  The simplest of relaxation processes are   single and double spin flip  processes considered by Overhauser \cite{Overhauser}. There was shown that the  most effective type is the first one originating from spin-current interaction which is the coupling between the magnetic moment of an electron and the magnetic field produced by the translational motion of another one.  The derivation presented in Ref.20 yields the relaxation rate which is many orders of magnitude smaller than the relaxation rate $\Gamma_{\bf q}$  of the order of several Kelvin found in UGe$_2$\cite{Huxley}. 
   So, the explanation of magnetic relaxation  in terms of itinerant nature of ferromagnetism in uranium compounds is represented unsolved.
   
   An interpretation of magnetostatic properties of uranium compounds  in frame  of itinerant electrons approach also looks doubtful.
Magnetic susceptibility of single UGe$_2$ crystals has been measured by Menovsky et al \cite{Menovsky} ( for the more recent results see  papers \cite {Sakon,Troc}). The easy axis magnetization  at zero temperature was found 1.43 $\mu_B/$f.u. that  in case of itinerant ferromagnetism corresponds to completely polarized single electron band.
On the other hand the  neutron scattering measurements of magnetic form factor \cite{Kernavanois} shows that:
(i) the shape of its $q$ dependence  is not distinguishable from the wave vector dependences of the  form factors of free U$^{3+}$ or U$^{4+}$ ions, (ii)
practically whole magnetic moment  both in paramagnetic and in ferromagnetic states concentrated at uranium atoms \cite{footnote} and (iii) its  low temperature value at $q\to 0$ coincides with magnetization measured by magnetometer with accuracy of the order 1 percent. 

   The configuration of localized $5f^2$ electrons of each atom of UGe$_2$ in paramagnetic state mostly consists of superposition of three quasidoublets and three singlets  arising from the state with fixed value of total momentum $J=4$ split by the crystal field.\cite{Troc}  The temperature decrease causes the change in probabilities of  populations of crystal field states revealing itself in temperature dependence of  magnetic moment. The quasidegenerate ground state formed by the lower quasidoublet allows the system to order magnetically with the ordered moment of $\sim1.5\mu_B$  twicely smaller than  Curie-Weiss moment deduced from susceptibility above the Curie temperature.

   The itinerant electron subsystem formed by $7s$, $6d$ and partly $5f$ electrons is also present  providing about $0.02\mu_B$ long range magnetic correlations
as demonstrated by muon spin relaxation measurements \cite{Yaouanc,Sakarya}. 

All mentioned observations  as well the theoretical treatment \cite{Troc} unequivocally point on the local nature of UGe$_2$ ferromagnetism.

The interaction between localized and itinerant electron subsystems   leads to the magnetization relaxation measured by neutron scattering  in paramagnetic and ferromagnetic state.
 This type of relaxation can be considered as analog of spin-lattice relaxation well known in physics of nuclear magnetic resonance \cite{Slichter}. In our case the magnetization created by the local moments of uranium atoms plays the role of "spin" subsystem, whereas the itinerant electrons present the "lattice" degrees of freedom absorbing and dissolving fluctuations of magnetization. According to this, we shall treat the  total magnetization almost completely determined by the local moments of uranium atoms as not conserved quantity. A deviation of magnetization from the equilibrium value  relaxes by transfer   to the itinerant electrons. 
 Unlike NMR relaxation determined by nucleus-electron magnetic moments  interaction the spin-lattice relaxation between the localized and conducting electrons is determined by spin-spin exchange processes and has no relativistic smallness typical for NMR relaxation.

Here we propose the phenomenological description of critical dynamics
based on specific for strongly anisotropic ferromagnet uranium compounds property that  magnetization  supported by the moments located at uranium atoms  is not conserved quantity. 
To be more concrete we shall discuss mostly UGe$_2$.

Let us discuss first relaxation  above the Curie temperature.  The relaxation   rate of the order parameter fluctuation is determined by deviation of system free energy 
\begin{equation}
{\cal F}=\int dV\left (F_{h}+K_{ i j}\frac{\partial {M_\alpha}}{\partial x_i}
\frac{\partial {M_\alpha}}{\partial x_j}\right )
\end{equation}
from equilibrium.
UGe$_2$  crystallizes in the orthorhombic structure with
 magnetic ordering  along $a$ crystallographic direction, 
and the homogeneous part of the free energy density is 
\begin{equation}
F_h=\alpha_x(T)M_x^2+\alpha_yM_y^2+\alpha_zM^2_z,
\end{equation}
\begin{equation}
\alpha_x(T)=\alpha_{x0}\frac{T-T_c}{T_c},
\end{equation}
$ \alpha_y>0,~\alpha_z>0$,
 whereas   gradient energy in orthorhombic crystal  written in exchange approximation \cite{electr} is determined by three nonzero constants $K_{xx}, K_{yy},
K_{zz}$.  The coordinates $x,y,z$ correspond to the $a,b,c$ crystallographic directions.

 To describe  homogeneous  relaxation  
together with  diffusion we shall use set of 
 kinetic equations \cite{Landau} relating to each magnetization component
\begin{equation}
\frac{\partial M_\alpha}{\partial t} =-A_{\alpha\beta}\frac{\delta {\cal F}}{\delta M_\beta},
\label{kin}
\end{equation}
where the kinetic coefficient matrix has three nonzero elements
$A_{xx},~ A_{yy},~A_{zz}$. 
One can rewrite the above equations as
 \begin{equation}
\frac{\partial M_\alpha}{\partial t} + \nabla_i j_{\alpha i}=-\frac{M_\alpha}{\tau_\alpha},
\label{kine}
\end{equation}
where
$
\tau_x^{-1}=2A_{xx}\alpha_{x0}\frac{T-T_c}{T_c}$, 
$\tau_y^{-1}=2A_{yy}\alpha_y,~\tau_z^{-1}=2A_{zz}\alpha_z$ and there is no summation over the repeating indices in the right hand side of this equation.
The components of spin diffusion currents  are 
\begin{equation}
 j_{\alpha i} =-2A_{\alpha\beta}K_{ij}
\frac{\partial {M_\beta}}{\partial x_j}.
\end{equation} 

Measurements reported in the paper \cite{Huxley} with scattering wave vector ${\bf q}$  parallel to the crystal $a$ axis revealed no extra scattering relative to the background while for the ${\bf q}$ parallel to the c-axis (${\bf q}\parallel\hat z$) a strongly temperature dependent contribution was found.
 The treatment similar  to that was used to get the  diffusion scattering function \cite{foot} given by Eqs.(\ref{d}),(\ref{diff})
yields 
\begin{equation}
S_{xx}(q_z,\omega)=\frac{2\omega\chi_{xx}(q_z)}{1-\exp(-\frac{\omega}{T})}\frac { \Gamma_{q_zx}}{\omega^2+\Gamma_{q_zx}^2},
\label{strfactor}
\end{equation}
and the same structure expressions for $S_{yy}(q_z,\omega)$ and $S_{zz}(q_z,\omega)$ correlators.
The corresponding widths of quasi-elastic scattering are
\begin{equation}
\Gamma_{q_zx}=2A_{xx}\left (\alpha_x(T)+K_{zz}q_z^2\right),
\label{Gammapara}
\end{equation}
$
\Gamma_{q_zy}=2A_{yy}(\alpha_{y}+K_{zz}q_z^2)$, and
$
\Gamma_{q_zz}=2A_{zz}(\alpha_{z}+K_{zz}q_z^2).$
The correlator $S_{xx}(q_z,\omega)$ having a form characteristic of critical magnetic scattering contributes the main part in differential cross section of scattering.
As one can see 
$\Gamma_{q_zx}$
does not vanish as $q_z\to 0$ for temperatures different from $T_c$ in correspondence with the results reported in the paper \cite{Huxley}.  
This property is the consequence of the  relaxation mechanism specific for ferromagnetic  uranium compounds where magnetization created by the moments located at uranium atoms.

Below Curie temperature in the ferromagnetic state the deviation magnetization from equilibrium value $M=M(T)$  is  $(M_x-M,M_y,M_z)$. The homogeneous part of free energy density of magnetic fluctuation is
\begin{equation}
F_h=2|\alpha_{x}(T)|(M_x-M)^2+\alpha_yM_y^2+\alpha_zM^2_z.
\end{equation}
 One can write kinetic equation similar to (\ref{kine}) only for the magnetization component parallel to ferromagnetic ordering
\begin{equation}
\frac{\partial (M_x-M)}{\partial t} + \nabla_i j_{x i}=-\frac{M_x-M}{\tau_x},
\label{kine1}
\end{equation}
with the same expression for the diffusion current as in paramagnet state.
Dynamics of perpendicular to  equilibrium magnetization components of magnetization is described by linearized Landau-Lifshitz-Gilbert equations
\cite{electr,LL,G}
\begin{eqnarray}
\frac{1}{\gamma}\frac{\partial (M_y+aM_z)}{\partial t}=-H_zM_z+h_z(t),\nonumber\\
\frac{1}{\gamma}\frac{\partial (M_y-aM_z)}{\partial t}=H_yM_y-h_y(t).
\end{eqnarray}
Here $\gamma$ is gyromagnetic ratio, $a$ is dimensionless damping parameter, $H_{y}=M(K_{ij}q_iq_j+|\alpha_x|+\alpha_{y}),~
H_{z}=M(K_{ij}q_iq_j+|\alpha_x|+\alpha_{z})$ are the components of "effective field" \cite{electr}, $h_{y}(t),~h_{z}(t)$  are the components of time dependent transverse external field. This set of equations determines the spin-wave spectrum  which has particular simple form in the absence of damping $\omega=\gamma\sqrt{H_yH_z}$.

These equations also determine the $({\bf q},\omega)$ dependences of $yy$ and $zz$ components of magnetic susceptibilities. In low frequency limit they are frequency independent and pure real : $\chi_{yy}=H_{y}^{-1},~\chi_{zz}=H_{z}^{-1}$. The latter means that according to the fluctuation-dissipation theorem they do not make a contribution to the corresponding components of dynamical structure factor determining the cross-section of neutron scattering. Thus, at $T<T_c$ the structure factor is given by the same formula  (\ref{strfactor})
as in paramagnet state, but the width of quasi-elastic scattering  now is given by
\begin{equation}
\Gamma_{q_zx}=2A_{xx}\left (2|\alpha_x(T)|+K_{zz}q_z^2\right).
\label{Gammaferro}
\end{equation}

The equations (\ref{Gammapara}) and (\ref{Gammaferro}) are  the main results of the paper. 
The line width of quasielastic neutron scattering near the Curie temperature proves to be  linear function of $T-T_c$. The  absolute value of the derivative $|d\Gamma_{q_zx}/dT|$ in ferromagnetic region is roughly twice as large as the corresponding derivative in paramagnetic region. The dependence of the wave vector $q_z$ is parabolic. All of these findings are  in qualitative correspondence with the experimental observations reported in the paper \cite{Huxley} (see Fig. 4 in this paper). 

Still there is one inconsistency with experimental findings.
Namely, there has been observed \cite{Huxley} (see Fig. 4d) that the product $\Gamma_{\bf q}\chi({\bf q})$ is temperature independent above Curie temperature but reveals the fast drop below $T_c$ contrary to the behavior described by the present phenomenological theory where this product is temperature independent both above and below $T_c$.  Measured   susceptibility $\chi({\bf q})$ decreasing with temperature proves to be  much faster than it is
 in accordance with mean field theory.

{\it Conclusion.} The totality of experimental observations point on the local nature of magnetism  in uranium ferromagnetic superconductors. The interaction between localized and itinerant electron subsystems gives rise to  specific mechanism  of magnetization relaxation similar to "spin-lattice" relaxation known in physics of nuclear magnetic resonance. This relaxation determined by exchange spin-spin coupling is much faster than NMR relaxation  supported by much weaker  interaction between electron and nuclei magnetic moments.
We developed a phenomenological description of quasi-elastic magnetic relaxation 
based on specific for heavy fermionic  ferromagnet uranium compounds property that  magnetization  supported by the moments located at uranium atoms  is not conserved quantity. As result the line width of quasi elastic neutron scattering at $q\to 0$ acquires nonvanishing value at all temperatures besides the Curie temperature.  The treatment  is the simple application of general description of critical relaxation proposed by Landau and Khalatnikov for a  case of nonconserving order parameter \cite{Landau}. 
The main message of the paper is that the nonconservation of magnetization in ferromagnetic superconducting uranium compounds points on the local character of magnetization  relaxing to equilibrium by the interaction with itinerant electron subsystem.

I acknowledge useful discussions with  M. E. Zhitomirsky. I am  also indebted to S.Raymond who presented me helpful possibility  to become acquainted with vast experimental literature
devoted to the critical dynamics in magnetic materials..


\begin{thebibliography}{20}
\bibitem{residual} Small residual linewidth $\Gamma_0\sim0.02~mev$  has been registered also in ferromagnet MnP. \cite{Yamada}
\bibitem{foot} It is worth to remind that the dynamical structure factor  is determined as $S(q,\omega)=2Re\tilde S(q,\omega+i\varepsilon)$, where $\tilde S(q,\omega)$ is the Laplace transformation for $\langle M(r,t)M(0,0)\rangle$  found from the diffusion equation (\ref{kine}). For the details of derivation see eg Ref.9.
\end{thebibliography}
\end{document}